# Unprecedented Quality Factors at Accelerating Gradients up to 45 MV/m in Niobium Superconducting Resonators via Low Temperature Nitrogen Infusion

A. Grassellino[1,a], A. Romanenko[1], Y. Trenikhina[1], M. Checchin[1], M. Martinello[1], O.S. Melnychuk[1], S. Chandrasekaran[1], D.A. Sergatskov[1], S. Posen[1], A.C. Crawford[1] , S. Aderhold[1], D. Bice[1]

[1]*Fermi National Accelerator Laboratory, Batavia, IL, 60510, USA*

## Abstract

We report the finding of new surface treatments that permit to manipulate the niobium resonator nitrogen content in the first few nanometers in a controlled way, and the resonator fundamental Mattis-Bardeen surface resistance and residual resistance accordingly. In particular, we find surface "infusion" conditions that systematically a) increase the quality factor of these 1.3 GHz superconducting radio frequency (SRF) bulk niobium resonators, up to very high gradients; b) increase the achievable accelerating gradient of the cavity compared to its own baseline with state-of-the-art surface processing. Cavities subject to the new surface process have larger than two times the state of the art Q at 2K for accelerating fields > 35 MV/m. Moreover, very high accelerating gradients ~ 45 MV/m are repeatedly reached, which correspond to peak magnetic surface fields of 190 mT, among the highest measured for bulk niobium cavities. These findings open the opportunity to tailor the surface impurity content distribution to maximize performance in Q and gradients, and have therefore very important implications on future performance and cost of SRF based accelerators. They also help deepen the understanding of the physics of the RF niobium cavity surface.

## Introduction

Superconducting radio frequency cavities are essential instruments for modern efficient particle accelerators and detectors, enabling progress and discoveries in different fields broadly ranging form particle physics, solid-state physics, material science, quantum computing, biology, medicine, industry and society. As the technology progresses towards enabling higher accelerating fields and/or higher efficiency at different RF fields amplitudes, new and unprecedented machines and detectors are enabled, allowing new scientific discoveries. Progress in achievable accelerating gradients and quality factors of niobium cavities has been driven by intense feedback work between empirical finds and the understanding of the physics of the first tens of nanometers of the niobium cavity surface, where impurity and defect content are known to play a dramatic role in quench fields, Mattis-Bardeen surface resistance, intrinsic residual resistance and even trapped flux induce residual resistance [1, 2, 3].

[a] annag@fnal.gov

In this paper we present the first set of measurements that demonstrate how to manipulate RF surface resistance and quench fields in a robustly controlled way via low temperature doping with nitrogen of the Nb cavity surface.

**Cavity Surface Preparation**

Several single cell bulk niobium cavities were used to conduct these studies; these cavities were fabricated out of RRR > 300 ATI Wah Chang 3 mm thick niobium material, by different cavity manufacturers as PAVAC Industries (identified as TE1PAV) and Advanced Energy systems (identified as TE1AES). Only one cavity was made out of large grain material, while the fine grain cavities ranged in grain size from 50 to 500 microns, as some of the material showed significant grain growth as the cavities were treated at high temperature. The treatment history of the bulk niobium of these cavities involves barrel polishing and/or electro-polishing as final surface finish. Previous studies [4] had proven that good RF performance and contamination free surfaces could be obtained if cavities are heat treated in a high vacuum high T furnace by taking the precaution of using protective niobium caps and niobium foil on the cavity flanges, as illustrated in figure 1. This is a crucial step for the successful surface preparation, as contaminants as titanium can reach the cavity surface and cause unwanted RF losses if caps -that allow gases in and out, but prevent line of sight path- are not used. The cavities are first ultrasonically cleaned and caps are installed in class 100 clean rooms. These protective caps and foils are buffered chemically polished and ultrasonically rinsed prior to every use. The cavities are then transported to the furnace capped directly through a class 10000 area. Installation sequence of the caps and foils to protect the inner surface from titanium is shown in figure 1.

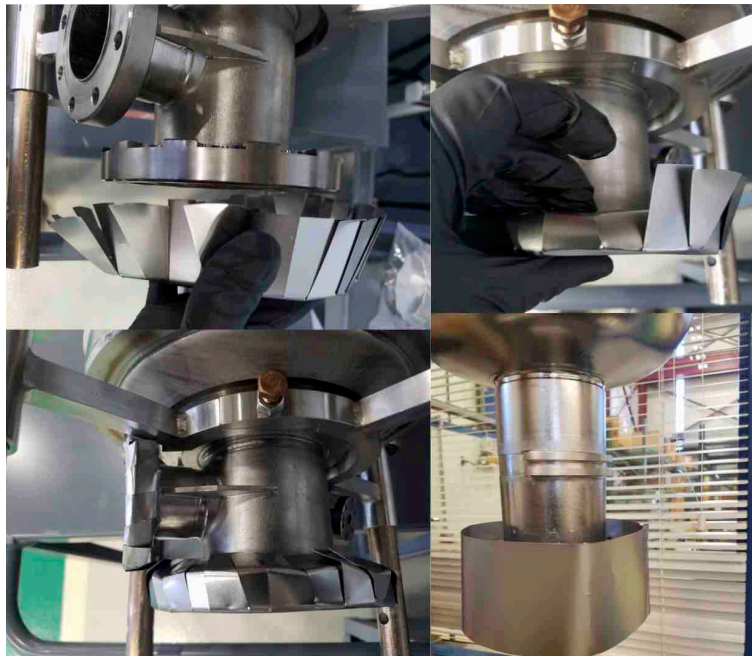

Fig.1 Cavity preparation with protective caps before heat treatment in high temperature high vacuum furnace.

The key point for these studies is that the surface preparation procedure will not involve any electropolishing post furnace treatment, so that we can study the effect of manipulation of impurity content on nanometric level. Therefore, it is important to ensure surface cleanliness prior to the bake, to avoid "baking in" any potential field emitters. The heat treatment is performed in a high temperature high vacuum furnace and involves several steps. The first step is always a hold at 800°C for 3 hours in high vacuum. This step is used for hydrogen degassing, but also for ensuring to break the native $Nb_2O_5$ oxide and disperse any oxygen coming from it in the bulk. At that point, the surface is naked and clean, and impurities can be introduced at low temperature for studying the effect of nanometer sized impurity layers on the surface resistance. The 800°C can also serve more in general as a surface reset for the same cavity treated multiple times, since 800°C 3 hours will disperse also any nitrogen or other surface defects generated during the previous low temperature bake. This surface reset strategy brings the advantage of avoiding having to electropolish the cavity in between heat treatments, allowing for a true comparison of RF performance as a function of heat treatment and independent of surface morphology.

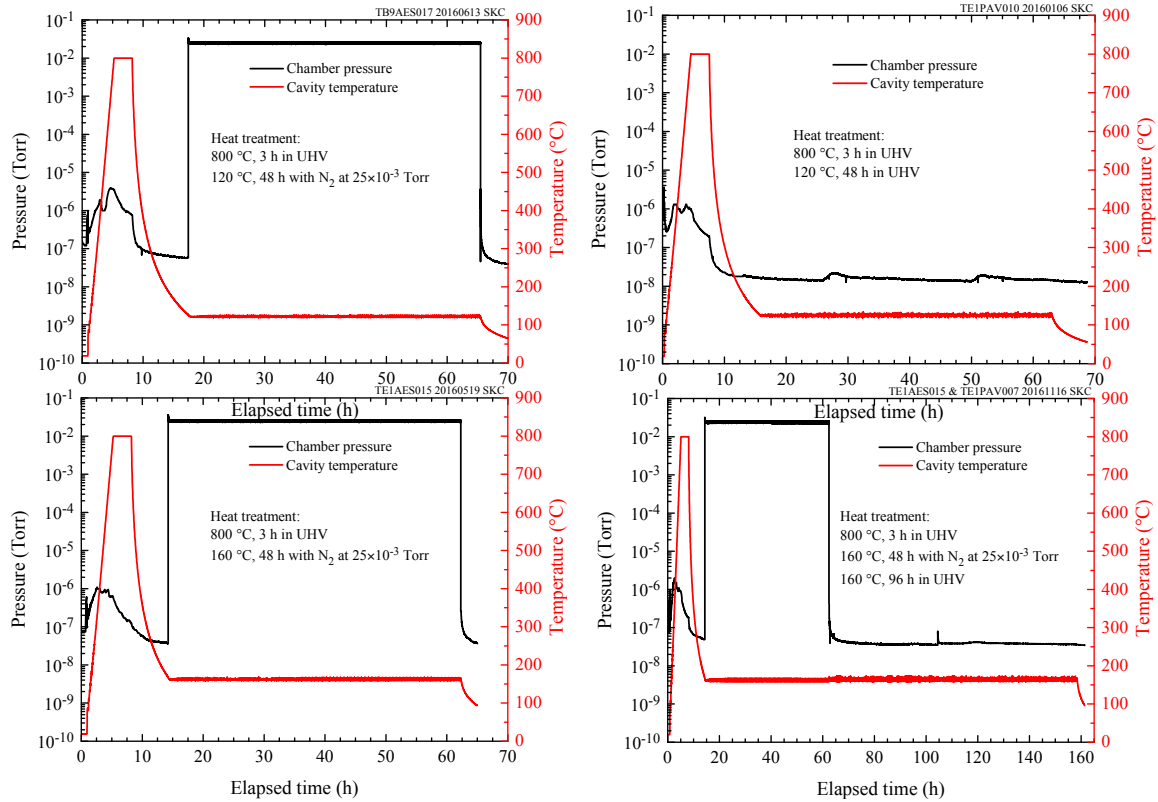

Fig.2 Temperature and pressure data for four different low temperature heat treatments.

After the three hours at 800°C, the temperature is lowered to a range of 120-200°C at which point the cryopumps are shut down, and nitrogen is injected in the furnace at a pressure of ~ 25 mTorr. Pressure is maintained via roughing pumps. For some of the treatments we did not inject nitrogen during the second step, for others we included a third step of hold at same low T but without nitrogen, as it will be explained more in detail in the following paragraphs. Four different examples of furnace treatments are shown in figure 2. Following the heat treatment, the cavity is transported to class 10 clean rooms, high pressure water rinsed and assembled with RF hardware, evacuated and sealed ready for RF vertical test.

## Experimental Results

### Nitrogen versus no nitrogen at 120°C: a cure for the high field Q-slope

The studies began with a comparison of cavities treated with 800°C 3 hours in HV, followed by 120°C for 48 hours without nitrogen versus 120°C for 48 hours with 25 mTorr of nitrogen, as shown in figure 2 a and b. The treatment with 120°C with no nitrogen was already studied by Ciovati [5] to attempt to better understand the role of oxygen in the elimination of the high field Q-slope (HFQS) typically obtained with the standard 120°C bake (starting from regular oxidized Nb surface). The idea behind this experiment is that if oxygen (or presence of the oxide on the surface) does play a role in the cure of the HFQS, then baking at 120°C for 48 hours in vacuum -after 800°C breaks up the oxide and fully disperses the oxygen in the bulk - should generate a surface with high field Q slope. If HFQS would not appear, then the experiment would unequivocally rule out oxygen role in the HFQS cure. In the previous experiment the conclusion was that oxygen plays no role, as the results post HV 120°C bake had shown no HFQS up to quench, which was reached ~30 MV/m (~120 mT in tesla shape), while HFQS onset in electropolished cavities systematically manifests at ~ 25 MV/m (~100 mT).

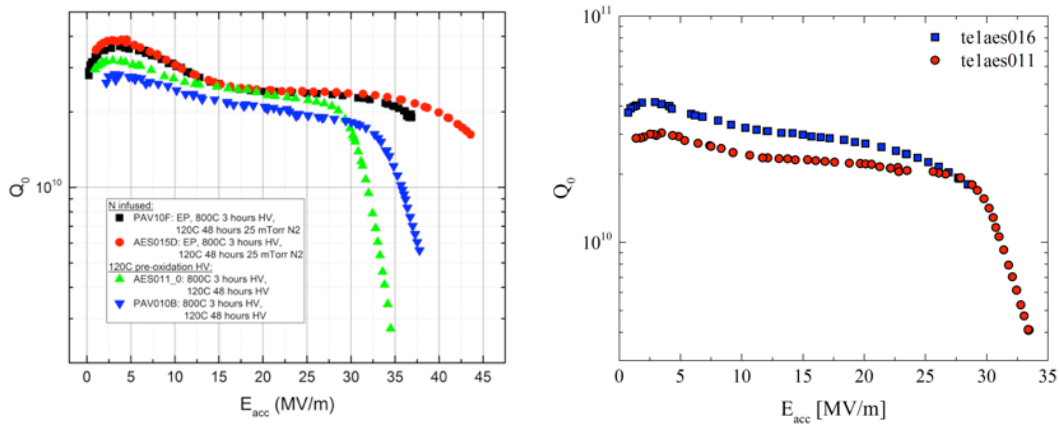

Fig.3 Left: effect of nitrogen injection versus HV bake at 120°C post 800°C 3 hours bake. Nitrogen injection helps fully eliminating the HFQS. Right – large grain versus fine grain treated with 800°C 3 hours in HV, followed by 120°C for 48 hours without nitrogen, showing both same onset of HFQS.

The results of our experiment are shown in figure 3. Interestingly, the first finding is that the vacuum 120°C bake post 800°C does not fully eliminate the HFQS, as it can be seen in fig. 3 for the cavities TE1PAV010, TE1AES011 and TE1AES016 (large grain), but rather shifts its onset up to ~ 30 MV/m, which can still be in line with the previous findings by Ciovati. One of the two cavities TE1PAV010 plus a different one TE1AES015 were then baked at 800°C for 3 hours, temperature was then lowered to

120°C with injection of 25 mTorr for 48 hours. As shown in figure 3 (left), the cavities did not exhibit HFQS up to quench fields of ~ 38 and 45 MV/m (~ 160 and 190 mT). The conclusion from this study is that a) 120°C bake post 800°C (so in absence of oxide) has some effect on HFQS (pushes the onset ~ 20% higher), but does not eliminate it completely; b) 120°C bake post 800°C (so in absence of oxide) but in atmosphere of nitrogen fully eliminates the HFQS. Interpretation of these findings will be discussed in a later section, but they unequivocally reveal that nitrogen does play a role in the elimination of the HFQS. It also plays a role in a systematic improvement of Q at mid and high field, as it is clear from figure 4 and 5. The origin of the improvement in Q stems from both an improvement in residual and BCS surface resistance, as a function of field, as it will be shown from the analysis of the decomposition [6] in the two components as a function of RF field, in figure 6.

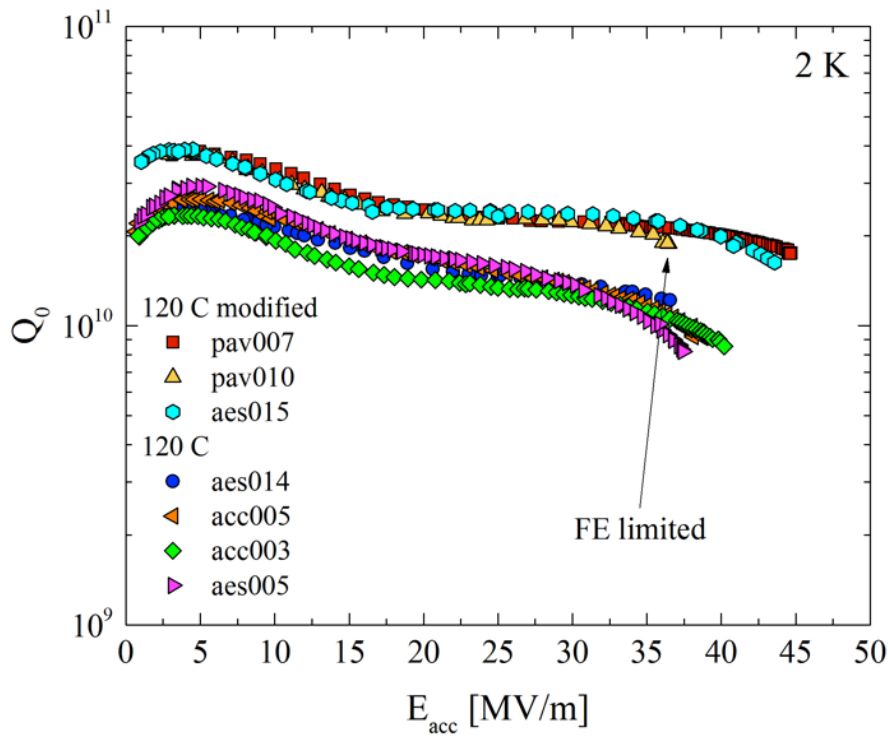

Fig.4 Effect of nitrogen infusion at 120°C versus standard 120°C bake post oxidation.

The 120°C for 48 hours in nitrogen pressure was applied to three different cavities. Figure 4 shows clearly the systematic improvement in Q and even accelerating gradients compared to cavities baked with the "standard" 120°C (post oxidation). This is even more strongly proved in the direct comparison shown in figure 5 where the same cavity is sequentially: a) treated at 800°C for 5 days, no EP after furnace treatment b) standard 120°C baked (post oxidation) c) 800°C baked for 3 hours followed by 120°C in the furnace with 25 mTorr of nitrogen. The sequence for identical morphological surface (no EP in between heat treatments) reveals a clear trend of improvement in Q and accelerating gradients. The 800°C produces the same results as a regular EP surface – large field independent BCS surface resistance ~ 14 nanoOhms (2K), and presence of

HFQS at 25 MV/m onset. The standard 120°C bake cures the HFQS, lowers the BCS surface resistance at low field but makes it highly field dependent, and increases the residual resistance and its field dependence. The 120°C with nitrogen infusion improves simultaneously the residual resistance and the BCS surface resistance. A comparison of the field dependence of BCS and residual surface resistance for the three different 120°C bake conditions are shown in figure 6.

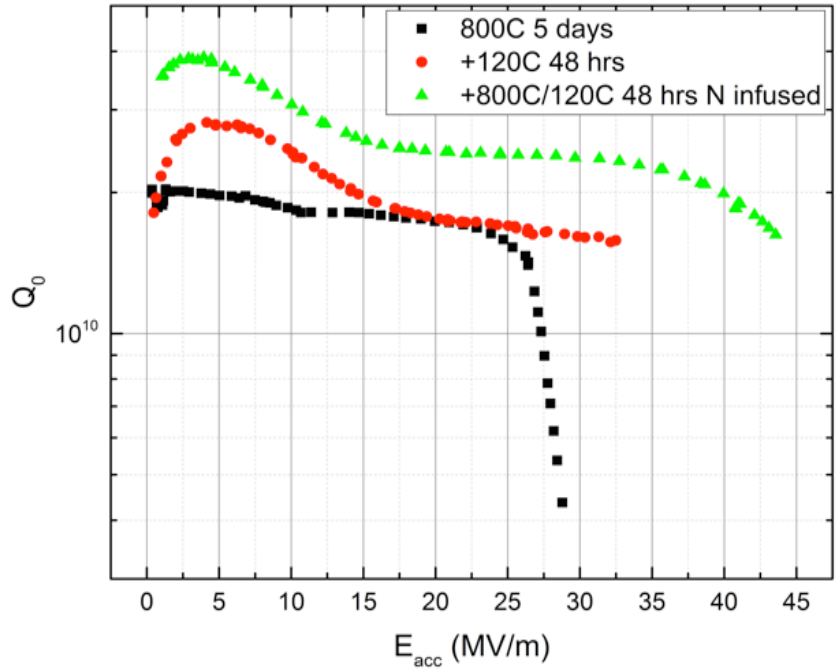

Fig.5 Cavity performance results for same cavity subject to different heat treatments. No electropolishing is done in between treatments for direct comparison of effect of surface impurities on same surface topology.

As it can be clearly seen from figure 6 comparison, the improvement in Q with the "oxide free" treatments (120°C in furnace directly after 800°C) stems from a reduction of the field dependent residual resistance of about a factor of two. An additional gain is coming from the reduction of the BCS surface resistance especially for the nitrogen infused treatments. So infusing nitrogen at 120°C helps a) eliminating the HFQS (field dependence of residual resistance in fig 6b below) and further lowering of the BCS surface resistance (fig. 6a). It is interesting to note that the field dependent BCS surface resistance increases with field, reaches a maximum and then starts decreasing with field. For the post oxidation 120°C the peak is reached at ~ 20 MV/m, while interestingly for the nitrogen infused 120°C bake, the peak moves to the left ~ 16 MV/m.

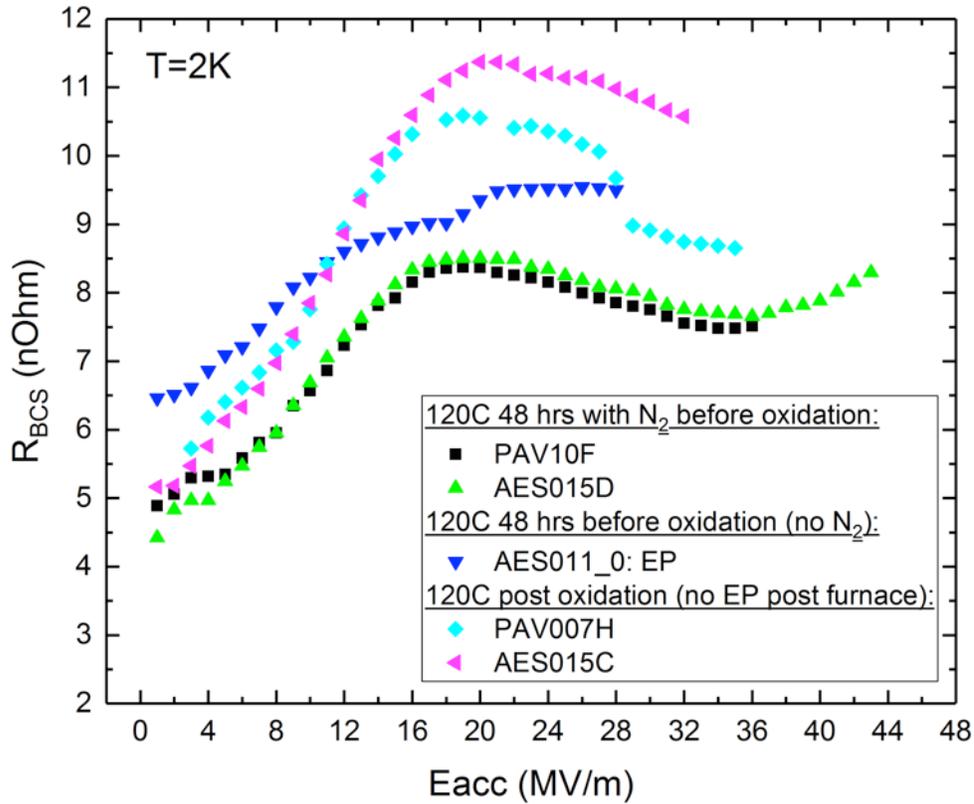

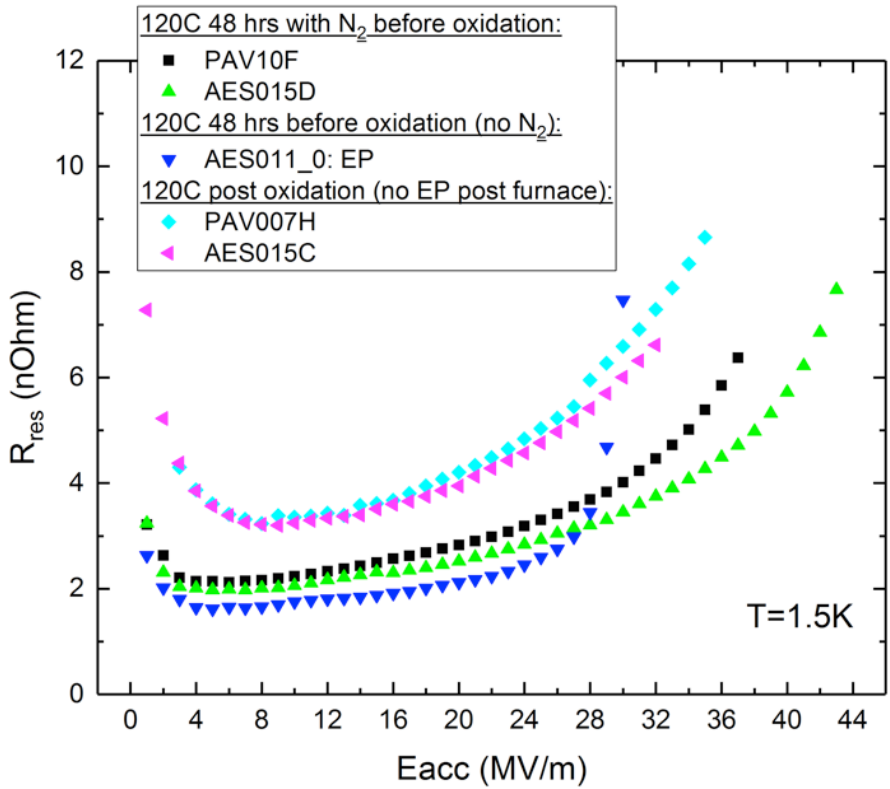

Fig.6 Comparison of decomposition in BCS and residual RF surface resistance as a function of field for standard 120°C bake, N infused 120°C bake, oxide free 120°C bake.

## N infusion for longer time and higher temperatures: the $R_{BCS}$ field dependence reversal

The effect of the low temperature nitrogen treatment becomes even more pronounced as time duration at 120°C is increased or as temperature is increased to 160°C and above, showing the typical anti Q-slope behavior of nitrogen doped cavities. This is consistent with nitrogen diffusing deeper in the surface and will be discussed in a later section. In fig. 7 we show a comparison for four different 120°C conditions: a) for 48 hours in high vacuum post 800°C 3 hours in HV; b) for 48 hours with 25 mTorr of nitrogen, post 800°C 3 hours in HV; c) for 96 hours with 25 mTorr of nitrogen, post 800°C 3 hours in HV d) standard 120°C bake post oxidation. It is interesting to see how the Q curve (fig.7) and the BCS surface resistance (fig.8) change dramatically with only doubling the time in nitrogen at 120°C, which would seem to indicate that the impurity that causes the antislope behavior is diffusing on the order of tens of nanometers. A premature Q-slope starting 13 MV/m is also found, reminiscent of HFQS behavior, or of the behavior of overdoped cavities.

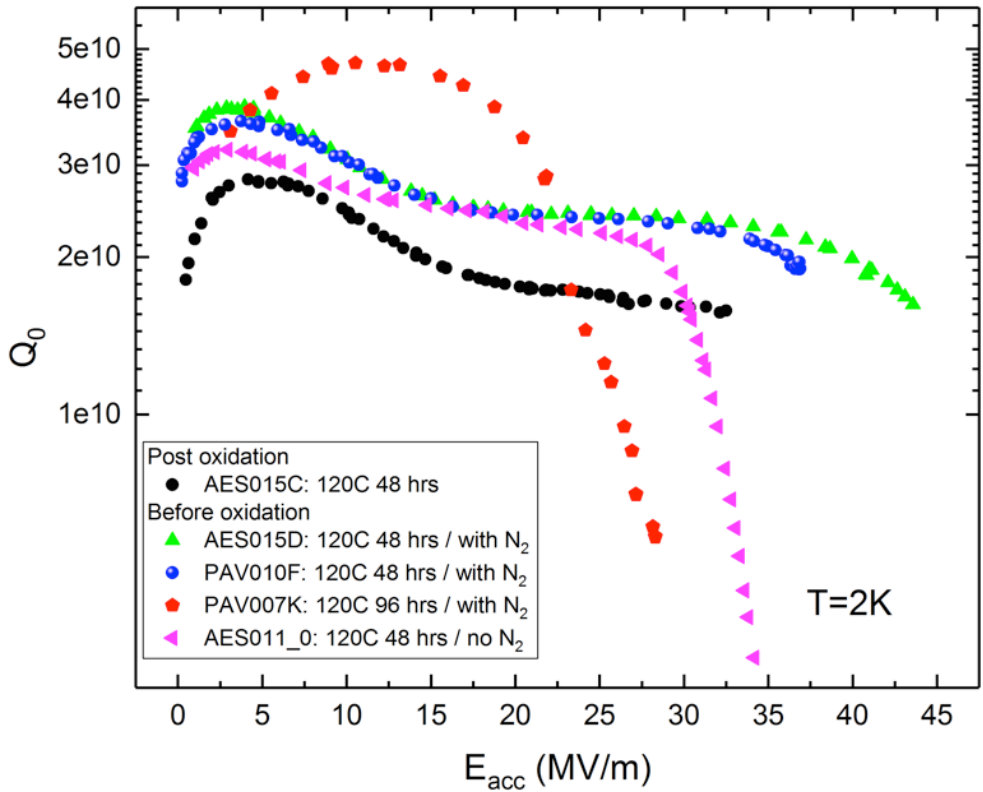

Fig.7 Comparison of Q curves for 120°C bake under different conditions – standard post oxidation, with and without nitrogen exposure for 48 hours, 96 hours.

Figure 9 shows another interesting comparison: for exact same cavity, identical surface topology, but only different surface impurity content as dictated by the sequence of heat treatments (where the 800°C 3 hours in HV each time resets the surface state), we can see that 160°C for 48 hours with nitrogen causes the reversal of the medium field Q-slope. This is confirmed by the surface resistance decomposition (fig 11) showing the typical reversal of the BCS resistance, which value-wise appears as typical doped (close to somewhat underdoped as it decreases only from 7 to ~ 6 nanoOhms) cavities. However, a Q-slope appears at higher fields. In appearance it seems similar to HFQS with a higher onset, however from the decomposition analysis shown in figure 11 and 12, we notice something very interesting and new: the high field slope appears in the BCS surface resistance, rather than in the residual. The standard HFQS of regular EP cavities typically appears in the residual resistance [6]. Further studies will be done in the future with T-map to understand the pattern and therefore origin of these high field losses.

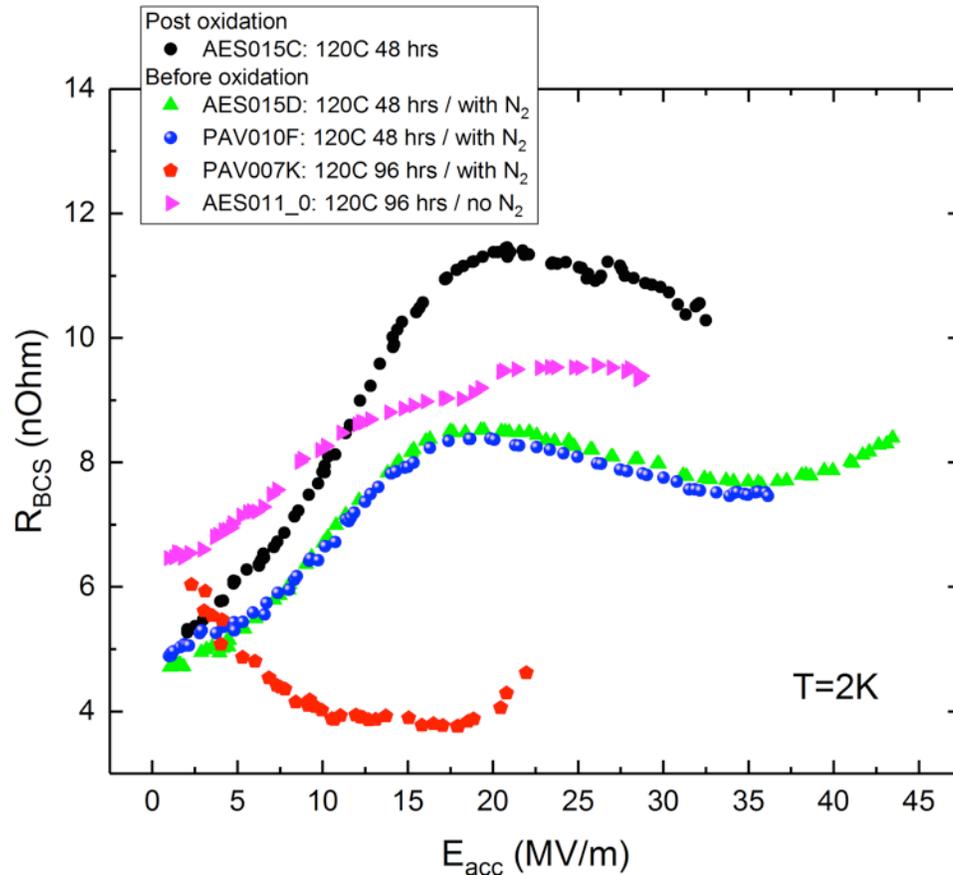

Fig. 8 Comparison of BCS surface resistance for 120°C bake under different conditions – standard post oxidation, with and without nitrogen exposure for 48 hours, 96 hours. The inverse field dependence appears simply doubling the exposure time to nitrogen at 120°C.

The same cavity was then baked at same T of 160°C for 48 hours with nitrogen but then followed by 96 hours with no nitrogen, with the idea to create a deeper and less steep nitrogen impurity profile. As it can be seen from figure 10 and 11, the BCS surface resistance now approaches fully that of ideally N doped cavities (decreasing from 7 to 4.5 nanoOhms), but the Q slope with 30 MV/m onset is again present and actually steeper than in the 160°C for 48 hours case. Another cavity, shown in figure 10, TE1AES011 was baked in an intermediate way - at 160°C for 48 hours with nitrogen followed by 48 hours with no nitrogen – and this cavity reached the highest fields without steep Q slope, reaching Q values >3e10 at 2K up to 32 MV/m, and above 2e10 up to 38 MV/m (quench limited in presence of x-rays). Then, when the same cavity was baked just 100°C higher - at 170°C for 48 hours with nitrogen followed by 48 hours with no nitrogen- the obtained quench field degraded significantly from 38 MV/m to ~28 MV/m no field emission present at 28 MV/m quench). As per Q values and surface resistance, in both cases impressive Q values were reached at mid-field approaching 5e10 at 2K, 20 MV/m. A difference can be found in BCS surface resistance: for the 160°C case, the BCS surface resistance approaches unprecedented values of <2 nanoOhms for 1.3GHz, 2K, >20 MV/m accelerating fields – essentially making Nb reach comparable to that of alternative material with larger critical T and SC gap as for example NbN, $Nb_3Sn$ or $MgB_2$.

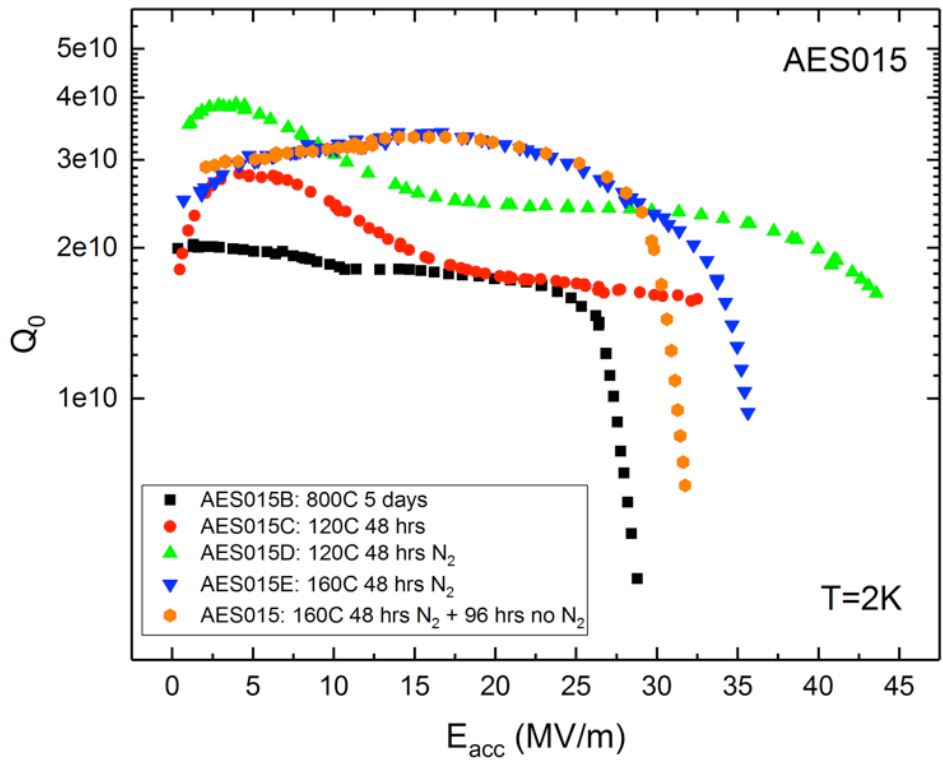

Fig. 9 Comparison of Q curves for same cavity and different heat treatment, highlighting the dramatic difference in RF performance for nanometric impurity changes at the surface.

A large grain cavity was also baked together with TE1AES011 with the 170°C recipe. As shown in figure 10, this cavity reached the highest Q out of all cavities, a record for 1.3 GHz cavities Q~6e10 at mid field, 2K. This is an interesting data point, as impurity diffusion could be assisted by grain boundaries and should be therefore different for fine grain and large grain cavities. Interestingly the large grain cavity resembles in RF performance the fine grain cavity baked for 96 hours at 120°C, with a very steep antislope and high Q achieved at mid field, but a premature onset of Q slope. The trapped flux sensitivity was also studied for these 170°C baked cavities and found to match the values of high temperature doped cavities treated with 2/6 min plus EP5 protocol [2].

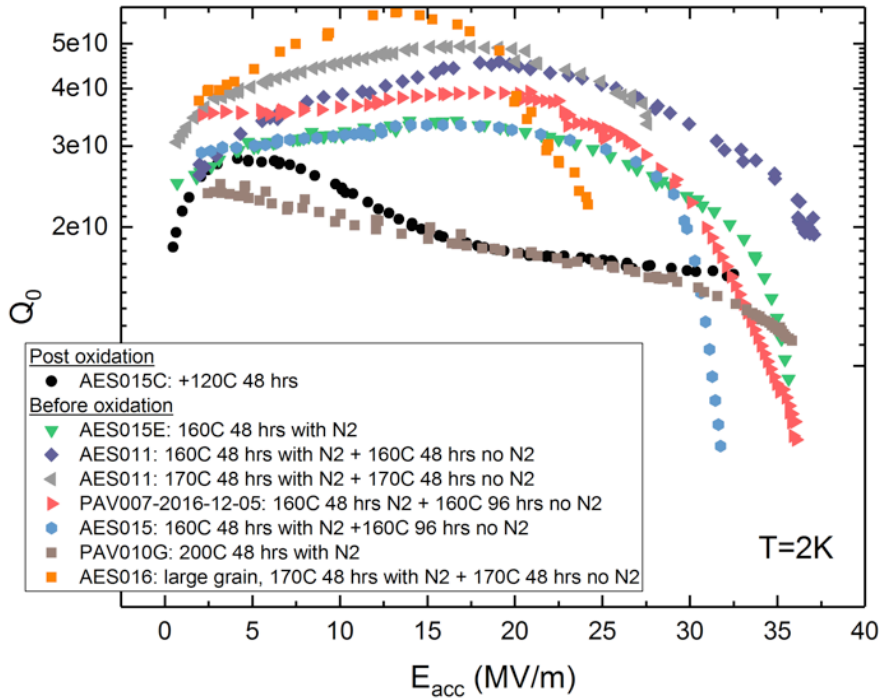

Fig. 10 Comparison of Q curves for all the different heat treatments in the range 160-200°C, referenced to standard 120°C bake post oxidation.

Another treatment was done at the higher temperature 200°C for 48 hours with nitrogen. As shown in fig. 10, the curve at 2K appeared like a regular EP curve, however from the decomposition in fig. 11 one could see that the BCS surface resistance was low, while the residual resistance was very large.

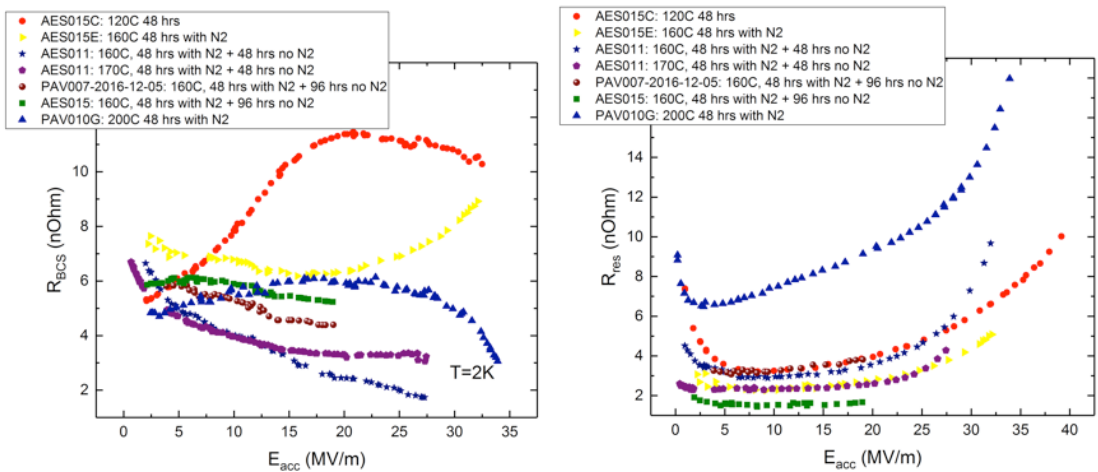

Fig. 11 Comparison of field dependence of BCS and residual surface resistance for all the different heat treatments in the range 160-200°C, referenced to standard 120°C bake post oxidation.

### Nitrogen versus no nitrogen injection at 160°C

During the gas injection period the cryopumps are shutoff and therefore the partial pressure of other gases as oxygen, carbon, hydrogen rises significantly. These gases will be also absorbed by the cavities. Therefore, to ensure that the effect of the reversal of the BCS surface resistance arises truly from nitrogen, two cavities were reset and baked with identical procedure, 800°C first, then 160°C for 48 hours, clean BCP'd caps and foils, with the difference that rather than injecting ultrapure nitrogen, ultrapure argon (25 mTorr) was used. The results are shown in figure 12, showing a comparison of same treatment (duration and temperatures) but different gas injected (nitrogen vs argon). The 2K curves of the argon treated cavities did not show anti-Q slope and actually performance were somewhat substandard compared to a regular EP cavity. The decomposition in BCS and residual resistance show a very large BCS resistance ~ 14 nanoOhms, about 3 times larger than that of doped cavities. A larger than typical residual resistance was also found, about three times of when the cavity was treated with nitrogen instead of argon. Differently from what claimed by the Cornell SRF group in [7], the results unequivocally prove that nitrogen is the key element generating the dramatic improvement in performance, in particular the peculiar reversal of the field dependence of the BCS surface resistance.

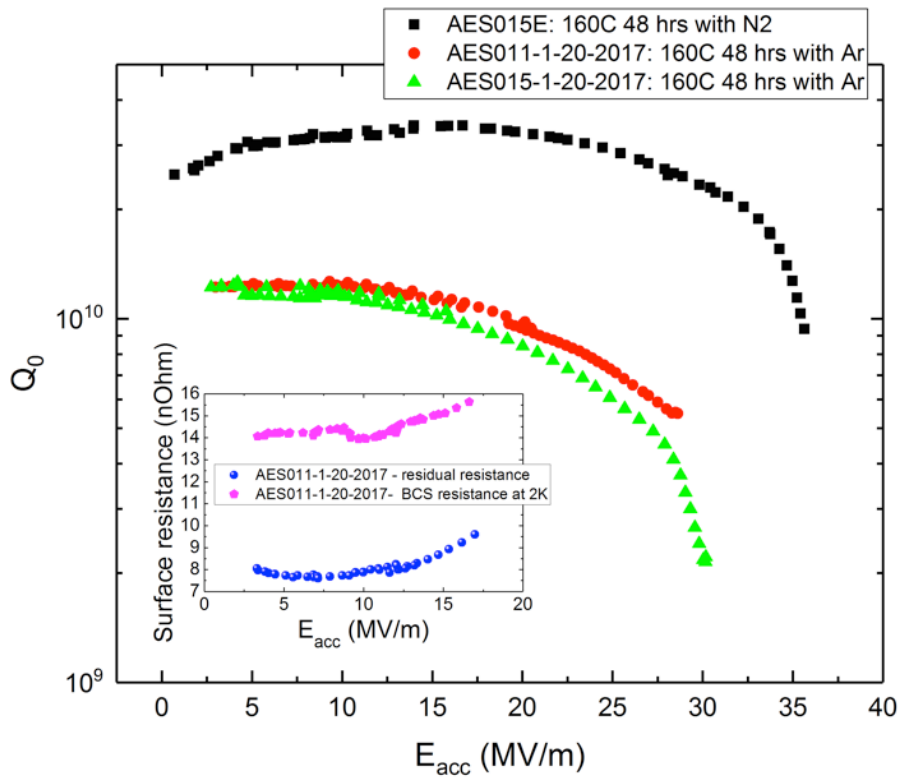

Fig. 12 Comparison of argon versus nitrogen treated cavities, injected for 48 hours at 160°C. Nitrogen is the key element causing the improvement in performance.

## Sample Analysis via SIMS, FIB/TEM and discussion

Samples treated together with the cavities at 120°C 48 hours and 160°C 48 hours versus a reference unbaked sample were analyzed with the TOF-SIMS technique at both IONTOF and FNAL. Two 160°C samples versus two reference and one 120°C samples were analyzed. Modern TOF-SIMS with optimized sputtering/analysis conditions has better than ppm level elemental sensitivity, as well as nanometer depth resolution. Results are shown in figure 13, where a relative comparison of non-baked, doped at 120°C and doped at 160°C can be seen. The x-axis is sputtering time, and for reference the $Nb_2O_5$ layer is plotted in each graph, to give a sense of the magnitude of depths – the RF layer where supercurrents flow is underneath the oxide and concentrations of impurities right under the oxide are of interest for RF performance – in the few to tens of nanometers. Some interesting conclusions can be drawn from these measurements. NbN signal (blue unbaked versus black 120°C versus red 160°C) shows that nitrogen concentration is increasing at the surface accordingly to the T increase of a factor of 10-20 compared to undoped background, right underneath the oxide layer.

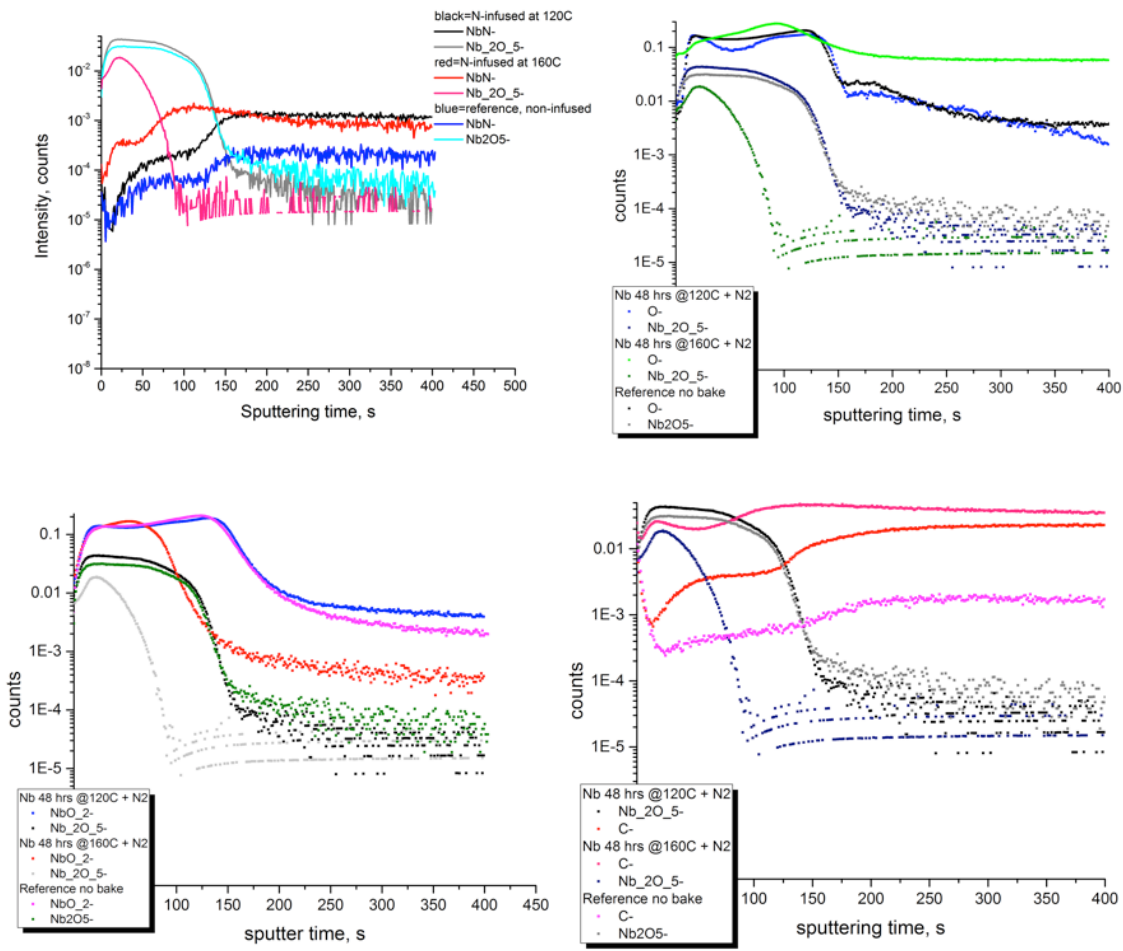

Fig. 13 TOF-SIMS profiles for nitrogen, carbon, oxygen, $Nb_2O_5$ and $NbO_2$ for unbaked, 120°C and 160°C doped samples.

The factor of 10-20 over nitrogen background is what has been shown to give the doping effect for the high temperature doping treatments [8], so this result is in agreement with the nitrogen doping effect as it was previously known. The difference stays in the fact that with the low T treatment the nitrogen enriched layer is just few-tens nanometers deep, versus microns for the high T treatments. The fact that 120°C (which does not lead to antislope) vs 160°C samples differ in nitrogen concentration just in the first few nanometers, gives yet again a sense of how important the first nanometers underneath the oxide layer are for RF performance, in agreement with previous HF studies in [2].

From the SIMS measurements we notice also some other interesting differences: the 160°C sample shows (systematically for both samples analyzed) a thinner oxide $Nb_2O_5$ (and also $NbO_2$ signal) of about half the depth compared to the undoped and 120°C case. This is a potentially very interesting finding, as it may be revealing that the improvement in performance via N doping/infusion may stem in part from the growth of a different, better quality oxide thanks to the presence of a certain needed concentration of nitrogen at the surface. This may be similar to what observed in semiconductors applications, where the growth of wet versus dry oxides plays a crucial role in performance of microelectronic devices. However, there is no difference in oxide nor oxygen content for the 120°C vs non baked sample, indicating that the removal of the high field Q slope is purely given by the presence of interstitial nitrogen underneath the $Nb_2O_5$ layer.

In terms of oxygen and carbon we can observe that both show a stronger signal increasing with the T. In particular oxygen shows no difference between undoped and 120°C, but stronger presence in 160°C as O- ion, however $NbO_2$ signal shows a lower tail compared to 120 and undoped. Carbon signal does increase accordingly to temperature. Combined with the cavity results, this tells us that interstitial carbon does not play a beneficial role on cavity performance as nitrogen does. Actually, cavity results shown in figure 12 may be showing that interstitial carbon in the absence of nitrogen may negatively impact cavity performance, as a much larger residual resistance than standard is found.

This could be explained by further studies of 160°C samples shown in figure 14, via TEM. As it can be seen, precipitates at grain boundaries are found, of the order ~ hundreds of nanometer size and depth. Further studies will be performed on the origin of these precipitates, which could likely be niobium carbides. So perhaps the presence of nitrogen helps binding carbon and reducing the size and number of carbide precipitates. In absence of nitrogen, carbides could be larger and cause the larger residual resistance observed. It is also possible that these precipitates found are responsible for the slope at high field found in 160°C nitrogen treated cavities, and that if one could really obtain a pure nitrogen diffusion and no other contaminants (eg no carbon), then cavity performance could be boosted further.

## Conclusions

The presented studies have shown for the first time a method to obtain a controlled layer of nanometric size enriched with nitrogen. The nitrogen infusion treatment at 120°C has proven to remove the high field Q slope and give high Q at very high gradients up to 45 MV/m. Increasing duration and temperature leads to the reversal of the BCS surface reistance and outstanding values of quality factors at mid to high fields up to 6e10 at 2K for 1.3 GHz cavities. Further studies are ongoing, exploring other temperatures and partial pressures of nitrogen, in feedback with SIMS studies, in search of a better optimum for cavity performance in terms of N enriched surface nano-layer.

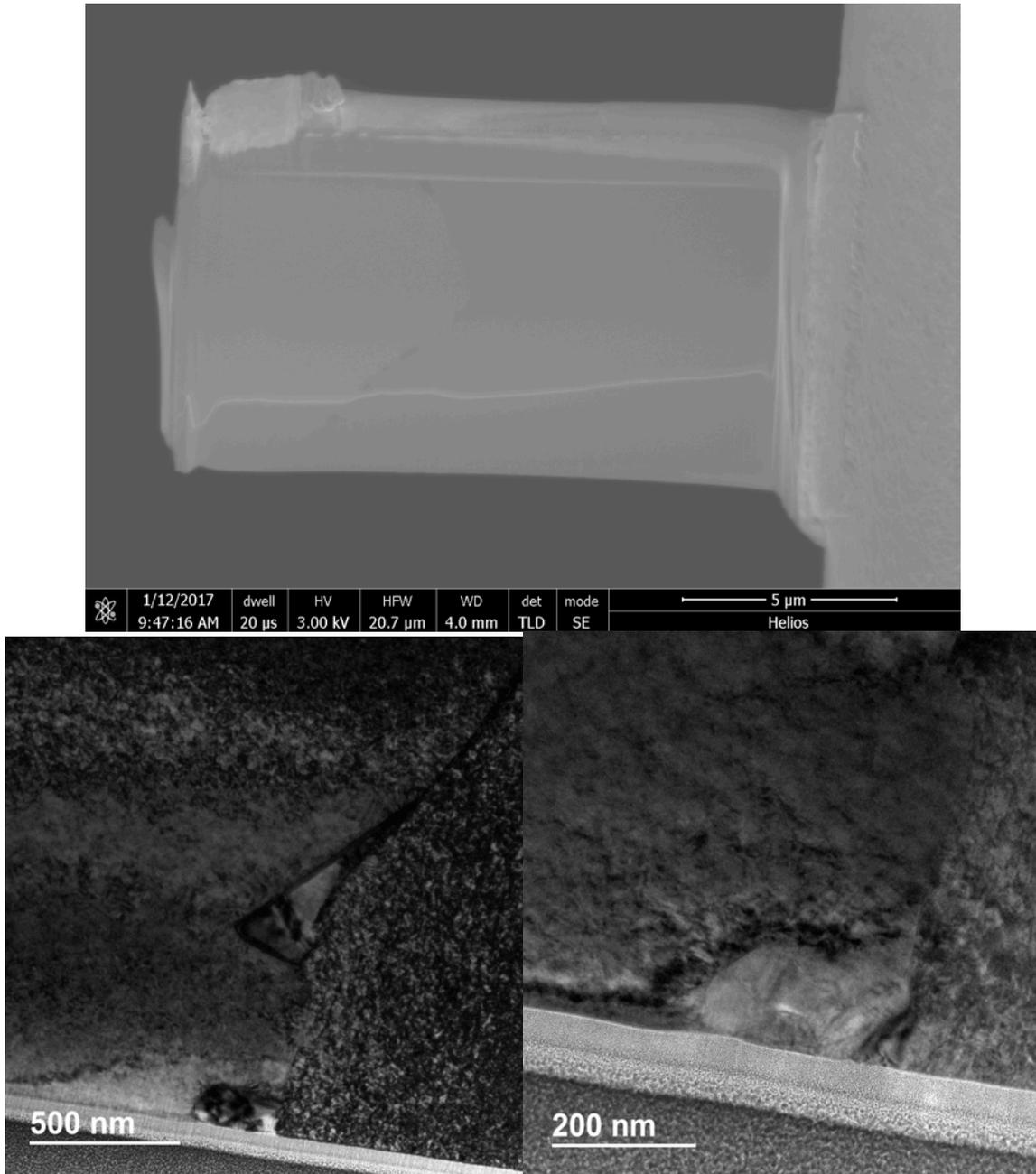

Fig. 14 TEM images of sample treated with nitrogen at 160°C. Precipitates at grain boundaries are found.